\documentclass[preprint,12pt]{aastex}
\usepackage{graphicx}

\newcommand{\microns}{\micron}

\newcommand{\etal}{{et~al.}}

\slugcomment{}
\shorttitle{The MIRI Focal Plane System}
\shortauthors{Ressler \etal}

\begin{document}

\title{The Mid-Infrared Instrument for the James Webb Space Telescope, VIII: The MIRI Focal Plane System}
\author{M. E. Ressler\altaffilmark{1}, K. G. Sukhatme\altaffilmark{1}, B. R. Franklin\altaffilmark{1}, J. C. Mahoney\altaffilmark{1},  M. P. Thelen\altaffilmark{1}, P. Bouchet\altaffilmark{2}, J. W. Colbert\altaffilmark{3}, Misty Cracraft\altaffilmark{4}, D. Dicken\altaffilmark{2}, R. Gastaud\altaffilmark{5}, G. B. Goodson\altaffilmark{1}, Paul Eccleston\altaffilmark{6}, V. Moreau\altaffilmark{2}, G. H.  Rieke\altaffilmark{7}, \& Analyn Schneider\altaffilmark{1}}

\altaffiltext{1}{Jet Propulsion Laboratory, California Institute of Technology, 4800 Oak Grove Drive, Pasadena, CA 91109, USA}
\email{Michael.E.Ressler@jpl.nasa.gov}
\altaffiltext{2}{Laboratoire AIM Paris-Saclay, CEA-IRFU/SAp, CNRS, Université Paris Diderot, F-91191 Gif-sur-Yvette, France}
\altaffiltext{3}{Spitzer Science Center, California Institute of Technology, Pasadena, CA 91125, USA}
\altaffiltext{4}{Space Telescope Science Institute, 3700 San Martin Drive, Baltimore, MD 21218, USA}
\altaffiltext{5}{DSM/Irfu/SEDI, CEA-Saclay, F-91191 Gif-sur-Yvette, France}
\altaffiltext{6}{RAL Space, STFC, Rutherford Appleton Lab., Harwell, Oxford, Didcot OX11 0QX, UK}
\altaffiltext{7}{Steward Observatory, University of Arizona, Tucson, AZ 85721, USA}

\begin{abstract}

  We describe the layout and unique features of the focal plane system for
  MIRI. We begin with the detector array and its readout integrated circuit (combining
the amplifier unit cells and the multiplexer), the electronics, and the steps by
which the data collection is controlled and the output signals are digitized and 
delivered to the JWST spececraft electronics system. We then discuss the operation of this
  MIRI data system, including detector readout patterns, operation of subarrays, and data
  formats. Finally, we summarize the performance of the system, including
  remaining anomalies that need to be corrected in the data pipeline.

\end{abstract}


\clearpage


\section{Detector System Overview}


The science potential of the James Webb Space Telescope (JWST) is derived from the 
rapid advances over the past three decades  in performance and size of infrared arrays. 
Some of the applications are discussed in Gardner et al. (2006).  The greatest gains with
JWST will be in the deep thermal infrared, where the high backgrounds on the ground have 
compromised the infrared array performance and previous cooled telescopes in space have had
small apertures and limited angular resolution; see Rieke et al. (2014a, hereafter Paper I).   

To implement fully the deep thermal infrared capabilities of JWST, the Mid-Infrared 
Instrument (MIRI) uses three arsenic-doped
impurity band conduction detector arrays, each of 1024$\times$1024 pixel
format with 25~$\mu$m pixel pitch. The performance expected is described in 
Glasse et al. (2014, hereafter Paper IX). These detectors have heritage to the
Si:As devices used in all three {\it Spitzer} instruments, but particularly
to the arrays in the Infrared Array Camera (IRAC) (Fazio et al. 2004; Hora et al. 2004). Like the IRAC arrays, the
MIRI devices were manufactured at Raytheon Vision Systems (RVS) of Goleta,
California. Both array types use a customized cryogenic readout process to
provide stable performance at low temperature and their detectors are
generally similar in terms of doping levels, layer thicknesses, and pixel
pitch.

\begin{figure}
\includegraphics[width=\textwidth]{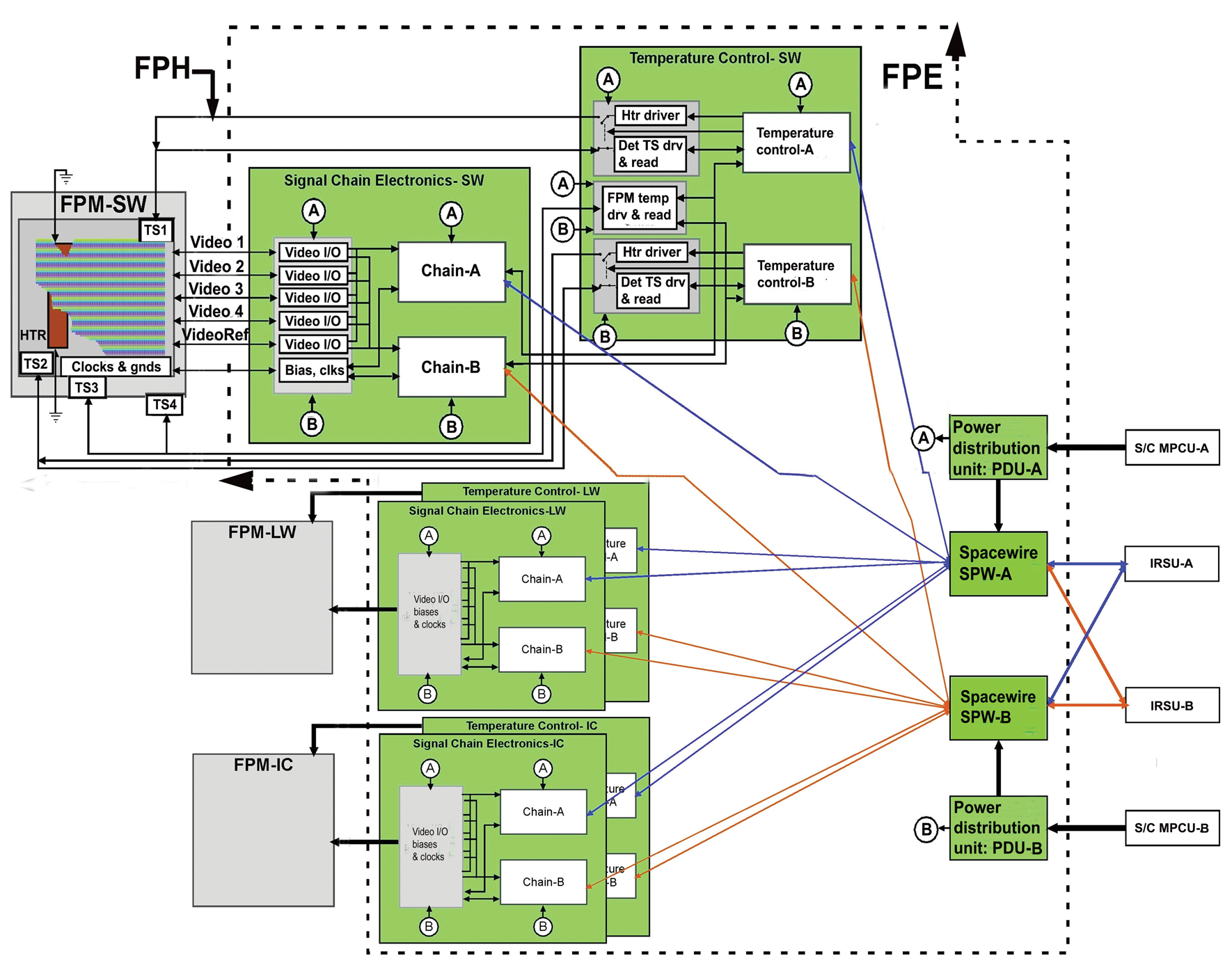}
\caption{Block diagram for the MIRI Focal Plane System.The uppermost of three identical input electronics 
trains is enlarged for clarity. \label{fig:fpsbd}}
\end{figure}

\begin{figure}
\includegraphics[width=\textwidth]{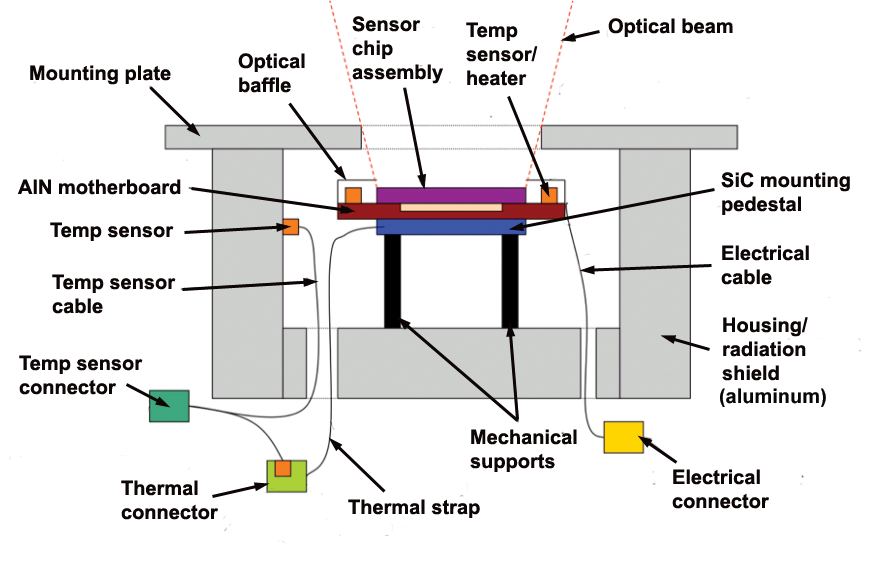}
\caption{Schematic diagram for the MIRI Focal Plane Modules.\label{fig:fpmbd}}
\end{figure}

\begin{figure}
\includegraphics[width=0.9\textwidth]{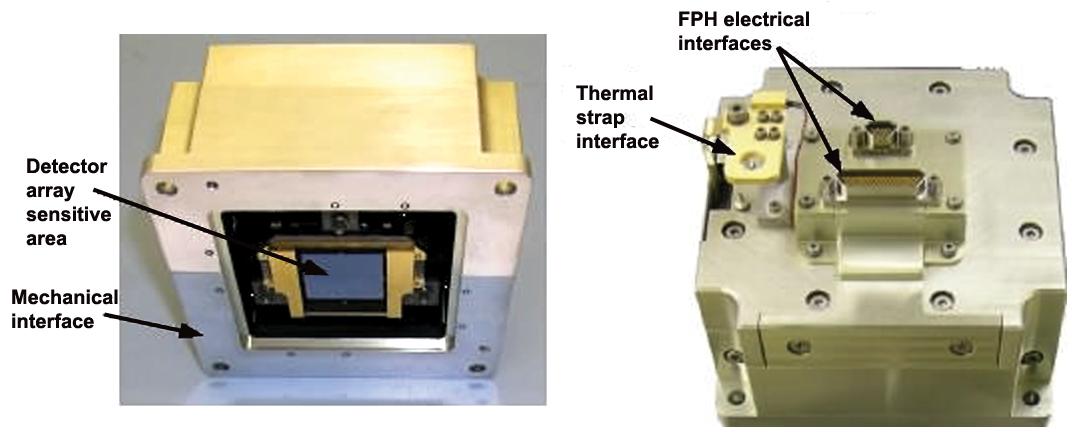}
\caption{Front and back views of a FPM. The aluminum housing is roughly 11 cm across;
the detector array sensitive area is 26mm square. \label{fig:fpm_frontback}}
\end{figure}

The MIRI detector system, or more formally the Focal Plane System (FPS), is
shown as a block diagram in Figure \ref{fig:fpsbd}. It is comprised of three
entities: the Focal Plane Modules (FPMs), the Focal Plane Electronics (FPE),
and the Focal Plane Harness (FPH). A FPM houses a detector array and locates
it at the relevant focal plane provided by the optical assembly (see Figures
\ref{fig:fpmbd} and \ref{fig:fpm_frontback}). There are three FPMs: one for
the imager, and one each for the shortwave and
longwave channels in the medium-resolution spectrometer (MRS).
The 4-m long FPH carries all electrical signals between the FPMs and the
FPE. The FPE consists of the control and readout electronics for the
detectors and also monitors and controls the temperature of each FPM to within 10 mK. Each
of the FPMs is driven by separate Signal Chain Electronics and Temperature
Control slices, with internal block redundancy (sides A and B). Our
discussion of this system begins with FPMs (\S 2) and FPE (\S 3), followed
by a description of the operation of the full FPS (\S 4) and a description
of the performance of the system (\S 5). Future work is previewed in \S 6.


\section{Focal Plane Module}

\subsection{FPM Design}

Each FPM (see Sukhatme et al. 2008) has a detector assembly (DA) and its
housing (Figure \ref{fig:fpmbd}). The DA includes the detector Sensor Chip
Assembly (SCA), heaters and temperature sensors, a fanout board, a
mechanical pedestal, an electrical ribbon cable, and connectors for the
signals to and from the SCA and for the temperature sensors (see Section
\ref{sec:da}). The FPM housing provides opto-mechanical alignment,
structural support, and thermal isolation for the DA (Figure
\ref{fig:fpm_frontback}). The housing has two auxiliary temperature
sensors to help monitor the thermal environment, and it is 20 mm thick, so
it also provides a major portion of the radiation shielding for the
detectors.

The DA mounting structure was designed to meet both the detector thermal
isolation requirements and provide the mechanical integrity to withstand
launch loads and the thermal expansion mismatch between the DA and the FPM
housing. The detector assemblies are supported within their housing by a
thermally isolating rod structure. This supporting structure keeps the
alignment of the sensitive surface of the detector array to within a 50 \micron{}
radius of the nominal detector position in X-Y, and within $\pm$20
\microns{} in Z/tip-tilt through the launch environment and from room
temperature down to an operating temperature of 6.7K or less.

For stray light reduction, serpentine, thermally insulating ports are
provided to pass the electrical cable and thermal strap through the housing
backplate. The cable is then attached to a thin aluminium bulkhead that
provides mechanical support for the connectors. The thermal strap is
supported by an insulating post and also connected to a thermal interface
plate. This plate is where the external heat strap is attached to provide
cooling for the SCA.

\subsection{Sensor Chip Assemblies\label{sec:sca}}

To manufacture the sensor chip assemblies for MIRI, the detector layers were
grown to MIRI-specific requirements, diced and patterned with indium bumps,
then bonded via matching indium bumps to the cryo-CMOS silicon readouts. The
resulting hybridized arrays are anti-reflection (AR) coated with one of two
possible single layer AR coatings, one optimized for 6 \microns{} and the
other for 16 \microns. Contrary to usual practice, to minimize interpixel capacitance the hybridized arrays
were not backfilled with epoxy; subsequent qualification testing proved the epoxy was unnecessary
for mechanical support purposes. More information about the detectors 
is provided in  Love et al. (2005) and Rieke et al. (2014b, hereafter Paper VII).

The readouts for these detectors are based on the cryogenic silicon circuit
process developed for IRAC and the far infrared detectors of MIPS on {\it
  Spitzer} and described in an early form by Lum et al. (1993). In this
approach, the circuitry is put on a thin surface layer of the silicon wafer; this
layer is grown on a degenerately-doped silicon substrate. This design brings
the ground plane through the substrate and close to the circuit even at very
low temperatures, improving the low-temperature electronic stability.

\begin{figure}
\includegraphics[width=\textwidth]{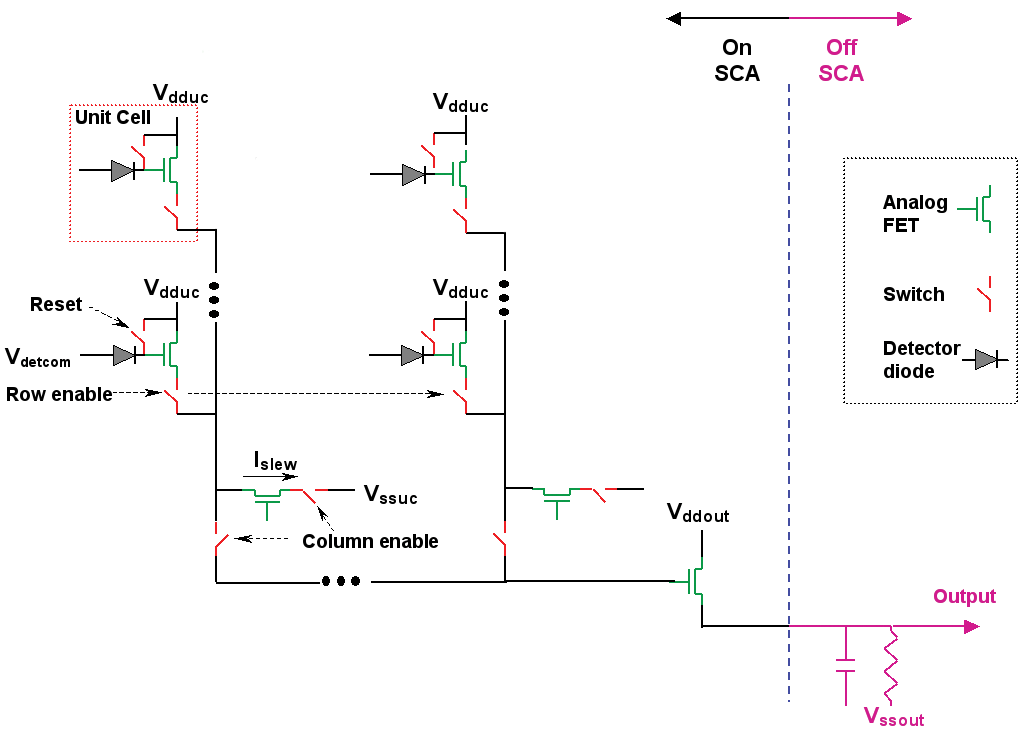}
\caption{Electrical schematic for the MIRI readout integrated circuit. Four representative 
unit cells are included along with the row and column multiplexing 
switching FETs and the output amplifier.The detectors themselves are
indicated as diodes because of their assymetric electrical characteristics. \label{fig:unit_cell}}
\end{figure}

The readout circuit is shown schematically in Figure
\ref{fig:unit_cell}, including four unit cells (each with an analog source follower FET and
a reset switch FET),  the row and column select transistors, 
and the output amplifier. An initial charge is placed
on the detector node capacitance through the V$_{dduc}$ supply when the
reset switch is closed to establish the detector bias voltage. 
The detector bias voltage is set by the difference between V$_{detcom}$
(applied to the transparent buried contact - see Paper VII) and V$_{dduc}$ (applied to the
indium bump contact). There is an additional $\sim 0.2$~V placed on the node
from clocking feedthrough so that the final applied bias voltage is
V$_{dduc}$ - V$_{detcom}$ + 0.2~V. After the switch is opened, photocurrent drains the
charge in proportion to the optical signal. The node voltage is buffered by
a source follower within the unit cell, then passed to the output source
follower/line driver through the row and column select switches. 

The completed hybridized SCA is mounted on an aluminum nitride motherboard, 
a material selected because its thermal contraction upon cooling approximately matches that of the
silicon detector array. The interface structure between the array and motherboard is designed to minimize the
residual stresses on the array.  

\subsection{Detector Assembly\label{sec:da}}

\begin{figure}
\includegraphics[width=0.9\textwidth]{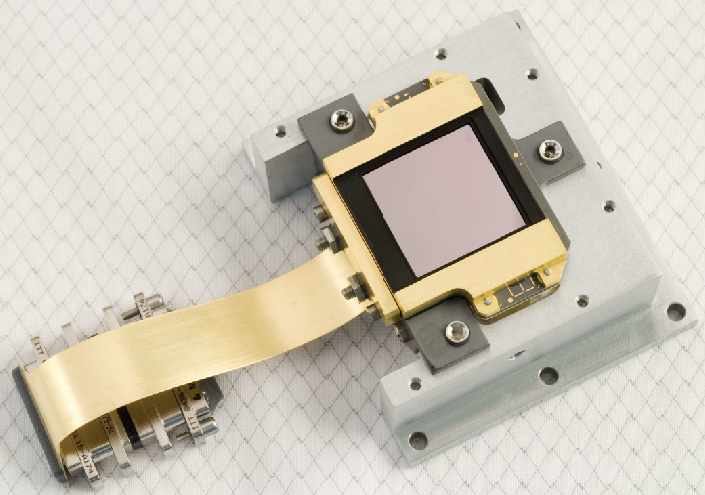}
\caption{Photograph of a MIRI Detector Assembly. The 26mm $\times$ 26mm detector sensitive
  surface is visible as the square in the center. Other components such as
  the temperature sensors, the filter resistors and capacitors, etc.\ are
  located on the fanout board under the gold-coated portions of the light
  shield. The plain aluminum structure is a shipping bracket and not part of
  the DA.\label{fig:dabd}}
\end{figure}

A picture of one of the MIRI detector assemblies is shown in Figure
\ref{fig:dabd}. The SCA is visible as the gray square in the center. The  multi-layer
aluminum nitride (AlN) fanout board is mostly covered by the gold-coated
light shield, but may be glimpsed at the lower right edge of the shield.
This, in turn, is mounted on a silicon carbide (SiC) mounting pedestal which
has six cylindrical bushings (3 visible) that serve as the mechanical
interface to the FPM housing.

A cable assembly that is electrically attached to the fanout board but
mechanically supported by the pedestal conveys electrical signals to and
from the SCA. Temperature sensors, heaters, and some R-C filters are mounted
on the motherboard, but are covered by the light shield. The electrical
interface is a 51-pin micro-D MDM connector. The thermal interface is
through a tab on the bottom of the SiC pedestal to which a copper thermal
strap is attached. The materials for the DA (AlN and SiC) were chosen to
have matching thermal expansion characteristics.

\section{Focal Plane Electronics}

\subsection{FPE Design}

Located in the room-temperature thermal zone behind the telescope,
(specifically, in the ISIM Electronics Compartment (IEC)(see Greenhouse et al. 2011  
for an overview of the ISIM and IEC)), the FPE contains
all of the control and readout electronics for the three detectors (Figure
\ref{fig:fpsbd}). Power is received from the JWST Integrated Science
Instrument Module (ISIM) and is converted to the requisite DC voltages
needed within the FPE by a Power-Distribution Unit (PDU) board (to the right in 
the figure). There are
primary and redundant PDUs within the FPE. Commands are received via a
SpaceWire communications interface board from the ISIM Instrument Control \&
Data Handling (ICDH) system and are interpreted by the FPE SpaceWire
interface cards (SPW, primary and redundant, also to the right). Valid commands are then
transmitted to the Signal Chain Electronics (SCE) boards (one per detector,
with primary and redundant sides on each board - to the left center, with the short wave 
(SW) detector channel enlarged for clarity), where the clock and bias
signals are generated  (video I/O, one signal chain for each of the four
array outputs and a fifth for the reference output) and sent down the 
harnesses to the detectors  (extreme left, 
SW channel enlarged for clarity). 


As directed by the clock signals, analog signal voltages from the detector are multiplexed and 
sent serially through the harnesses to be collected by the SCE video I/O 
boards, where they are amplified and digitized. The digital signals are packetized in the 
SPW boards (to the right) and then delivered to the ISIM Remote 
Services Unit (IRSU, extreme right) where they are stored for
later transmission to the ground. The SCE boards also collect telemetry reporting 
the detector control voltages and the board supply voltages to monitor the
health of the system. These are also sent back through the SPW boards,
though in separately identified packets.

The detector temperature control is provided by Temperature Control
Electronics (TCE) boards, again one per detector (center of the figure, enlarged for
the SW channel). The control is a
relatively standard proportional-integral-derivative (PID) feedback algorithm, 
although the differential term is
hard-coded to zero. Due to a flaw in the embedded heaters that are located
immediately under the detectors in the FPMs, we use the temperature sensors
located at opposite corners of the SCAs also as heaters since they are
simple resistive elements (though highly temperature sensitive!). The TCE
board drives power through the temperature sensor for 88\% of a 100 ms drive
cycle; for the other 12\%, the power is reduced to sensing levels and the
temperature is measured. When self-heating in the sensor is taken into
account, this provides a robust thermal control circuit, with drifts of less
than 1 mK (as monitored with the redundant sensor using ground support
electronics)\footnote{This temperature sensor performance supports controlling the temperature of 
the SCAs to 10 mK, peak-to-peak.}.

\subsection{Readout process}

\begin{figure}
\includegraphics[width=\textwidth]{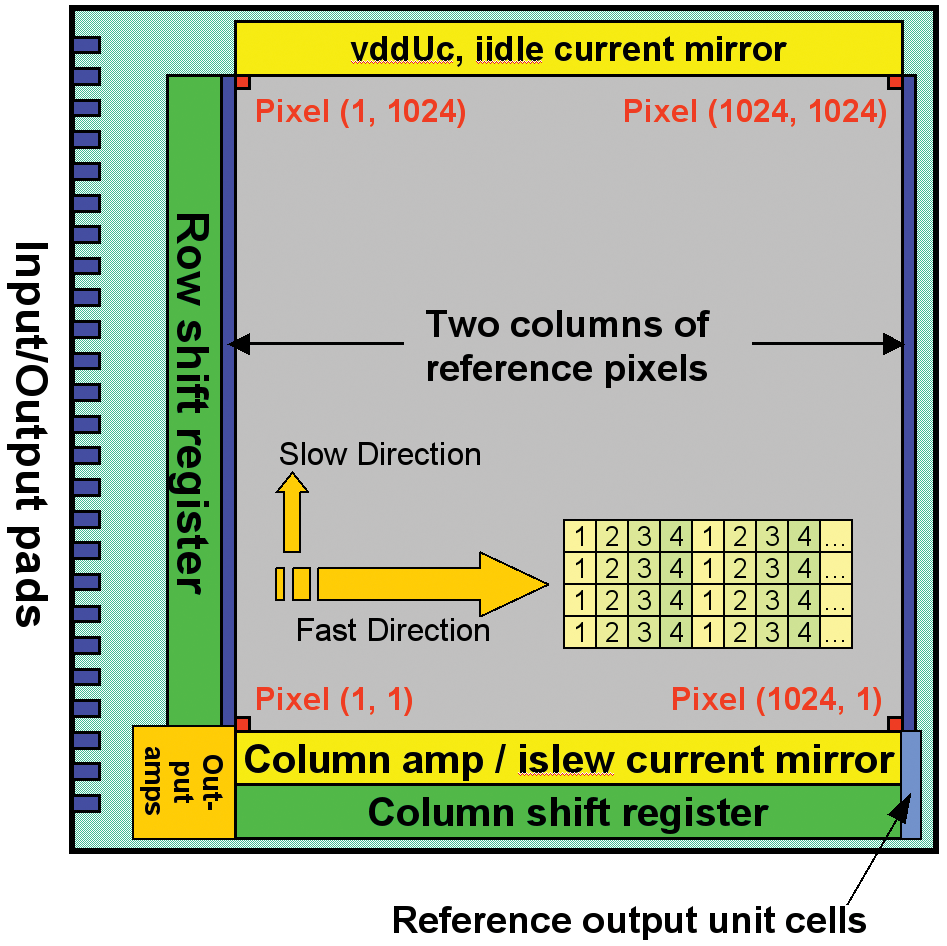}
\caption{Schematic of the readout multiplexer. The square area (grey) marked
  by the pixel corners corresponds to the light sensitive pixel region. The
  two narrow columns (blue) on the left and right of the light sensitive
  region identify the two sets of four columns of reference pixel locations.
  Row and column shift registers (dark green) border the pixels on the left
  and bottom and address pixel locations. The arrows and repeating 1234
  numbers show schematically the pattern by which the SCA is read out. The
  bond pads for the FPE signals are located on the left side of the
  array.\label{fig:mux}}
\end{figure}

A representation of the full MIRI readout is shown in Figure \ref{fig:mux}.
The SCA has a total of 1024$\times$1024 active pixels. There are four
additional ``reference pixels'' at the beginning of each row and four at the
end that are not connected to detectors, but in all other ways are treated
as light sensitive pixels. All these pixels (including the reference pixels)
are read out through 4 interleaved data outputs; each output presents
258$\times$1024 pixels to the signal chain electronics for processing. The
outputs are read simultaneously, so at a sampling rate of 10 $\mu$s per
pixel, it takes slightly less than 3 seconds to read out the full array.
Clock patterns control the row and column shift registers to address the
individual pixels for either destructive or non-destructive reading. Note
that all readout patterns start closest to the output amplifiers (lower
left hand corner) and this fact has been used to determine a preferred
orientation of the SCAs with respect to the various instrument focal planes.
There is an additional, fifth ``reference output'' that is a group of blind
pixels that are sampled continuously (and simultaneously with the 4 data
outputs) and that can be used for various engineering and data
quality-monitoring purposes. Since this signal also appears as a 
258$\times$1024-pixel data stream that is interleaved with the 4 data
channels, the SCA effectively presents a 1290$\times$1024-pixel ``image'' to
the outside world.

The pixel coordinate system starts with ``1'', not ``0''. Zero has
a special value to the readout shift registers, and so the first physical
pixel (actually the left-edge reference pixel) starts at column 1. The first
light sensitive pixel starts at column 5 and ends with column 1028; the last
right-edge reference pixel is column 1032. There are no reference pixels
along the top or bottom rows, so the rows are all active and range from 1 to
1024. \clearpage

\section{FPS Operation}

\subsection{Electronics Upgrade}

The Signal Chain Electronics (SCE) boards used for the flight instrument testing 
have a flaw that produced occasional corrupted science data frames. These boards 
have been redesigned to eliminate this problem, but with the requirement that they 
be pin-compatible with the original boards and therefore that they preserve the
basic operational characteristics of the system. Nonetheless, the new boards include 
a number of additional improvements. The boards are being switched at the end of 2014 and
the final flight data system for MIRI will be validated and calibrated during the third cryo-vacuum test 
of the ISIM in the second half of 2015. We describe the operation of this system below. 

\subsection{Individual Frames}

The FPS is controlled by commands received from the ICDH via the MIRI flight
software; it acts on them and returns science and telemetry data. All three
detectors are operated independently via the three SCE boards, but in a
similar fashion (i.e. there are no commands unique to individual arrays).
Observers will define their exposures from a palette of readout patterns
that have been predefined to support the full capability of MIRI. The
readout patterns for MIRI fall within the framework of the general
MULTIACCUM readout patterns\footnote{This readout pattern differs from that used with 
IRAC, in which multiple `Fowler' samples were obtained at the beginnings and ends 
of the integration ramps and only their differences were sent to the ground.} 
 adopted by the JWST mission so that all
instruments will have similar exposure interfaces.

The general scheme for reading out the SCA, referring to Figure
\ref{fig:mux}, starts with addressing row 1 and then reads left-to-right
through all 1032 columns (four at a time) before proceeding onto row 2, etc. until row 1024
is completed. Thus, the ``fast'' direction of the readout is across each row
and the ``slow'' direction is up the columns. The four outputs present four
pixels simultaneously to the electronics in a [1234] pixel pattern; this
block of four pixels repeats 258 times to complete an entire row of 1032
pixels. Recall that the first four and last four pixels of each row are
reference pixels. The 4th readout line, for example, is responsible for
every 4th column of data in the final image and will produce a ``jailbar''
pattern in the raw frames if its offset differs substantially from the other
outputs.

The pixels are reset by row pairs (i.e. 2 rows, 2064 pixels, at a time). For
example, row 1 will be read, then row 2 will be read, then they will be
reset together, then row 3 will be read, etc. The column and row shift
registers must be clocked through the entire SCA before returning to a given
odd-numbered row. This approach enables a final read immediately before
resetting the SCA, and thus captures the longest possible integration time.
The disadvantage of this approach is that one cannot read the values
immediately after reset, and thus the reset level (as in a traditional
correlated double sample) cannot be obtained.

\subsection{FASTMode and SLOWMode}

The electronics have the ability to sample an individual pixel
multiple times before moving on to the next pixel (sometimes referred to as
``dwell''). In ``FASTMode'', MIRI samples a pixel once during a single clock
cycle spent on that pixel. However, in ``SLOWMode'', ten samples of a pixel
are obtained and a subset of them can be averaged (e.g. four or eight) to
output a single result from the FPE to the IRSU, within an overall cadence
of $\sim$30 seconds. This approach can reduce MUX glow, for example, since
it is no longer necessary to read the pixel every $\sim$3 seconds as in
FASTmode. However, SLOWMode, because of the 30 second readout time, is most
appropriate for faint source and/or low background work.

\subsection{Clocking patterns}

\begin{figure}
\includegraphics[width=\textwidth]{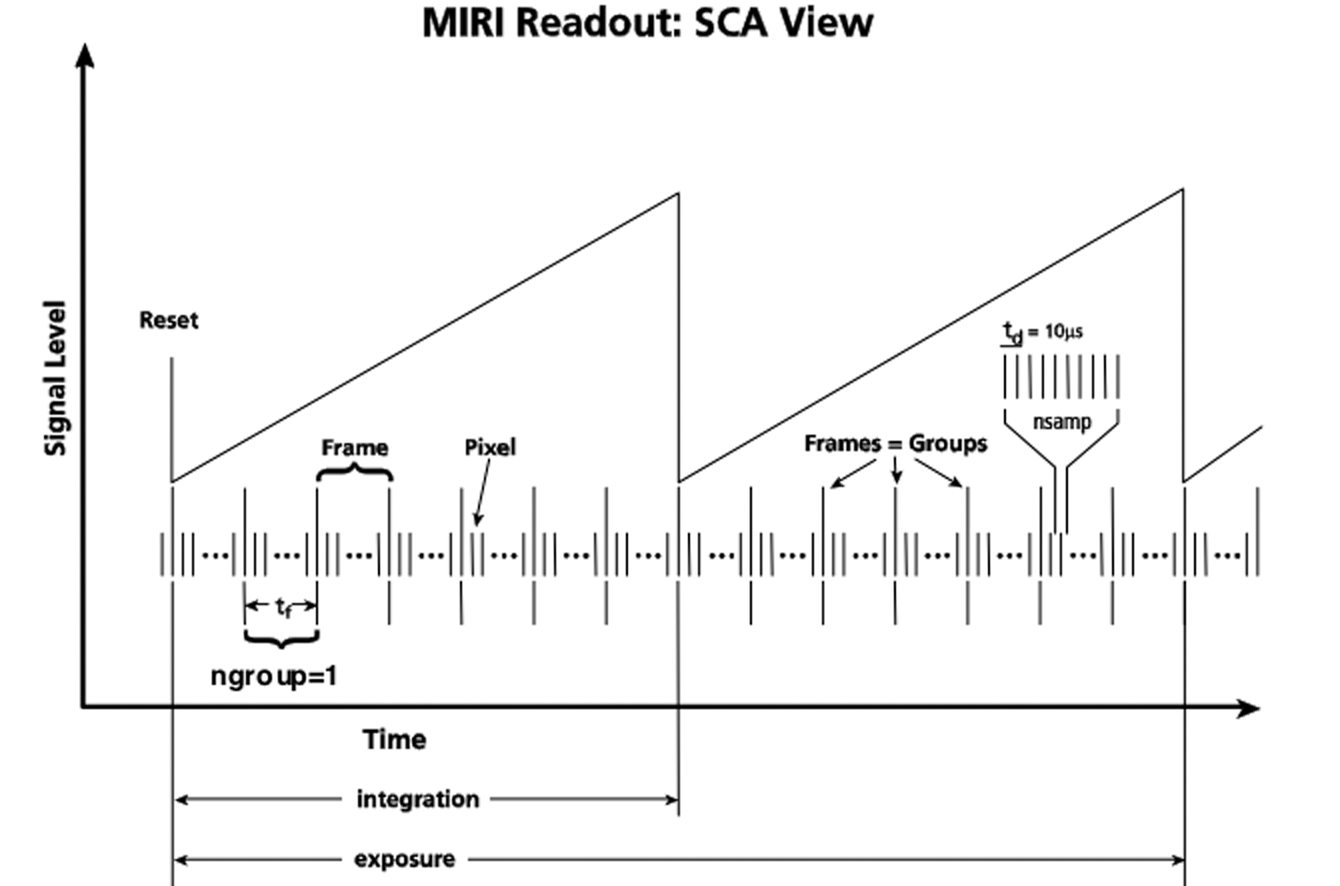}
\caption{Clocking patterns to read out the MIRI arrays. A representative integration ramp 
is included to help illustrate the definitions of key terms. \label{fig:clocking}}
\end{figure}

A schematic MIRI readout timing pattern is shown in Figure
\ref{fig:clocking}. In describing it, we adopt the detector lexicon of the
JWST Mission Operations Concept Document for reference.

The underlying approach is always to address the SCA at a constant rate
(whether exposing or not) to maintain the stability of the SCA properties,
in particular, the SCA temperature as well as slowly varying electrical
properties. The pixels of the SCAs will be continuously addressed at time
intervals of 10 $\mu$s, which is the time between pixel samples, $t_d$, and
which is set by the FPE master 100 kHz clock. The general MIRI timing
pattern is defined by only three of the MULTIACCUM parameters: 1) nsample,
the number of samples per pixel (for MIRI, this will either be 1 for
FASTMode, or 10 for SLOWMode), 2) ngroups, the number of groups during an
integration, where a group is the product of cycling through all the pixels,
and 3) nint, the number of integrations during an exposure, where
integration is defined as the time between resets. By definition, for MIRI
there is exactly one frame per group, so that ``frames'' and ``groups''
could be used interchangeably. The value of nsample (the READMODE parameter
given above) determines the time between frames, t1. The value of ngroup
determines the integration time, tint, as follows: tint = ngroup $\times$
t1. For example, 10 frames of FASTMode yield a tint = 10 $\times$ 2.775 =
27.75 seconds or $\sim$ 30 seconds. Note that with MIRI SCAs the delta time
between groups, t2, is zero. An exposure consists of one or more identical
integrations. The value for nint determines the exposure time as follows,
texp = nint $\times$ tint. For example, if we expose for 5 integrations with
a tint = 27.75 seconds, then texp = 138.75 seconds and during this exposure
time there were 5 resets of the array. There is no ``dead time'' between
frames or integrations, so the wall-clock time for an exposure is an
integral multiple of the single frame time. MIRI receives commands only at
an exposure boundary, so an exposure with its various parameters is the
lowest level of external commandability of MIRI.

One advantage of the MULTIACCUM readout mode is that cosmic rays can be
rejected using ground-based software that processes pixel samples taken
before and after a cosmic ray hit. MIRI can take maximum advantage of such
software because MIRI plans to download all its data to the ground for
processing, except in some high-data-volume cases. Rauscher et al. (2007)
studied the effects of cosmic rays on the imaging exposure times on the JWST
near-IR cameras, and these results are applicable to MIRI. For an
anticipated cosmic ray flux of 5 protons cm$^{-2}$ s$^{-1}$, they expect
that 20\% of the pixels will be affected in a 1000 second exposure.
However, the data pipeline is planned to identify cosmic ray hits within 
integration ramps and recover the data prior to and after them, to improve
efficiency. Regan \& Stockman (2001) discuss how MULTIACCUM type readout
facilitates integration times longer than 1000 seconds. The advantages of
this approach have been demonstrated on-orbit by the NICMOS instrument on
HST and the Spitzer/MIPS germanium detectors. We have demonstrated these
approaches for MIRI by testing various slope fitting algorithms described by
Fixen et al. (2000) and Offenberg et al. (2001) on data taken with flight
and flight-like MIRI sensors.

Thus, the integration time is determined not by how long it takes for a
specified fraction of pixels to be hit by cosmic rays; rather, it is set by
how long it takes for other noise sources to dominate. For broadband
imaging, the optimum integration time will typically be the time required to
become background limited. For MRS spectroscopy, the integration time should
be influenced by the time required to become dominated by shot noise on
detector dark current. Nonetheless, there will be residual noise from cosmic
rays that do not free many electrons, as well as an accumulated bias shift
from multiple hits. It may be desirable to reset the detectors periodically
to restore the bias voltage on them. Also, low frequency instability of the
detector readout and electronics may be the dominant noise in long
integrations.

\subsection{Data Format Including the Reference Output Line}

As described above, the MIRI SCAs have two types of reference signals to
assist with noise reduction: the left- and right-edge reference pixels, and
the reference output. The reference output is sampled simultaneously with
the four science data outputs, so that a full output frame is effectively
1290$\times$1024 pixels.
In addition to science frames (to be sent to the ground), the data are used
by the ICDH scripting engine to provide target acquisition and centering
information. To have a valid image, the scripts must skip every 5th pixel
(the reference output) as they ingest data.

\begin{figure}
\begin{center}
\includegraphics[height=2.7in]{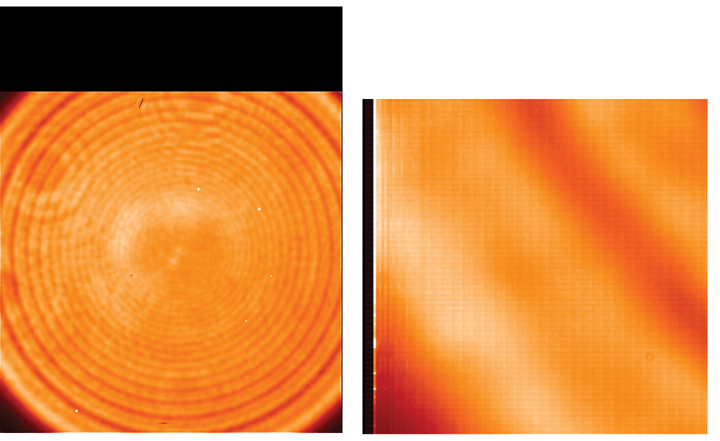}
\end{center}
\caption{Real data after it has been rearranged with the reference output
  data at the top. The reference pixels are along the side of the array as
  seen in the magnification at right.\label{fig:fpsdata}. The bullseye
  illumination is an artifact of the light source. These data were obtained at JPL using flight-clone electronics 
  but without the MIRI optical module. }
\end{figure}

The signal level in the reference output line may be substantially different
from the data output lines (especially in the case of high background
images), which will affect the compressibility of the data. To
improve this compressibility, the ICDH system rearranges the data stream
coming from the MIRI FPE so that all reference output pixels will be placed
at the end (top) of each frame as shown in Figure \ref{fig:fpsdata}, not in
every 5th column. This is the format that will be sent to the ground and
will appear in the ``Level 1'' (raw) FITS files.

\subsection{Subarrays}

The MIRI SCAs have the ability to read out partial frames through the
manipulation of the clocking patterns. This allows a reduction in the frame
time, and thus will allow integration times of substantially less than 3
seconds. For all subarrays, the subarray portion (the region-of-interest,
ROI) is read out and stored while the remaining rows of the array are
continually reset.

The subarray coordinates cannot be randomly accessed; one must step the row
and column shift registers from the origin (1,1) to the starting subarray
corner before proceeding. So at the beginning of the frame read, the first
two rows of the full array are accessed briefly and reset, then the 2nd
pair, etc. until the ROI is reached. For more efficient subarray operation, a “burst mode” clocks through the left-hand
columns at 5 times the normal speed.  In the first row of the ROI, the pixels
on the left that are not part of the ROI are clocked through (and not
digitized), after which the pixels that are part of the ROI are read. The
pixels to the right of the ROI are ignored by resetting the column shift
register to 0 (recall the special shift register definition) immediately
after the last ROI pixel. This pattern is repeated through all the rows
contained within the ROI. The rows after the ROI are stepped through quickly
and reset as were the rows before the ROI.

There are consequences to having to clock through the pixels on the left
side of the ROI. Suppose one wishes to observe with the SUB256
(256$\times$256 pixel) subarray that starts at pixel (413,51). The clocks
quickly sweep through the first 50 rows (70 $\mu$s per row or about 3.5 ms
total), but when one gets to row 51, 412 columns to the left of the ROI must
be skipped. If we could not quickly burst through them, 103 cycles would be
needed to get to the ROI (recall we access 4 columns at a time), then 64
more cycles are needed to read the pixels within the ROI itself (plus a few
cycles of overhead) for a total of 1.77 ms per row, or 453 ms for all the
rows containing the ROI. However, since we can choose to burst through the
412 columns at 5 times the normal rate, it takes only 21 10-$\mu$s periods
(103/5, rounded up) to get to the ROI, for a total of 0.96 ms per row or 246
ms for the area. The 718 rows after the ROI are also clocked at 70 $\mu$/s
per row for an additional 50 ms. The total frame time is the sum of these
three totals, 300 ms for the case where we use burst clocking vs 507 ms
where we do not. This compares to the 243 ms that are needed to read a same
sized subarray that starts in Column 1. Therefore, subarrays become slower
the farther they are from the left hand edge of the array, though this is
compensated somewhat by the use of burst clocking. This fact drove the
orientation of the imager array, since it is advantageous to have the
fastest subarrays located within the coronagraph.

\begin{figure}
\includegraphics[width=\textwidth]{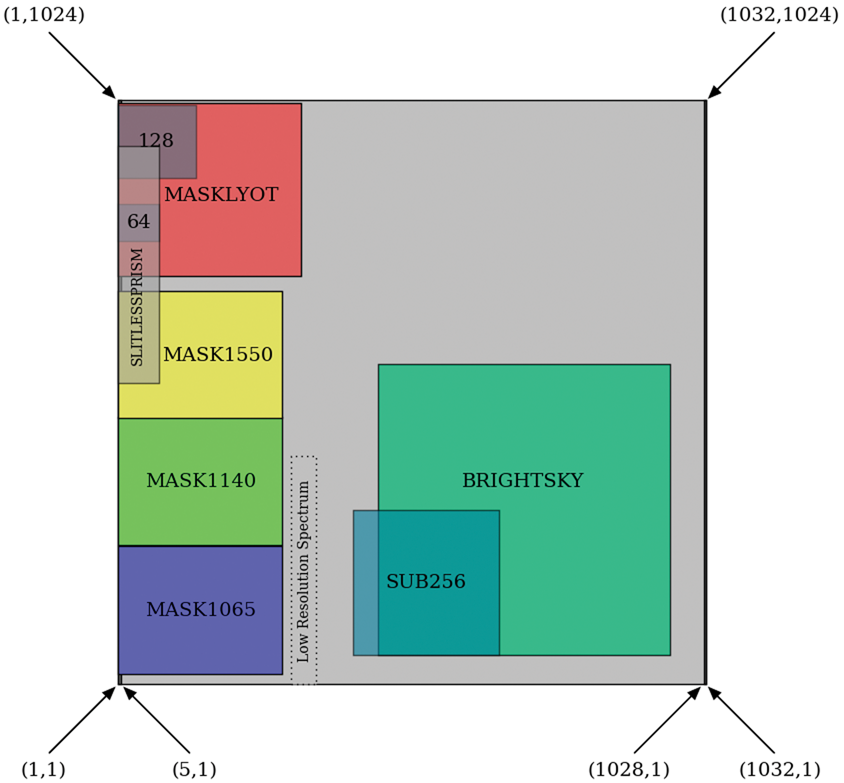}
\caption{Positions of subarrays in the imager field of
  view.\label{fig:subarrays}}
\end{figure}

As a result of the constraints, the arrangement of subarrays and the
resulting minimum integration times (and maximum fluxes) are complex. A
summary is provided in Table 1, with more discussion in Paper IX. Figure \ref{fig:subarrays} identifies the
proposed subarrays and their applications. To ensure consistent calibration
of subarray modes, only a few subarrays will be supported and the user will
have to select from these predefined versions. There are nine in total: 1.)
one for each of the four coronagraphic areas; 2.) one for high backgrounds;
3.) three for bright objects; and 4.) one for slitless LRS spectroscopy.
Subarray modes are only useful and provided for the MIRI imaging SCA, not
the MRS.

There are multiple considerations that led to the specific arrangement in
Figure \ref{fig:subarrays}. The total number of pixels contained within a 
subarray must be a multiple of
64, including the reference output values, to fit well into the ICDH architecture.
The coronagraph subarray locations and sizes are determined in part by the
final focal plane mask design; however, the actual size of the coronagraphic
subarray may be larger than its field of view to accommodate the multiple of
64 requirement on the subarray size. The coronagraph fields of view may be
used for both target acquisition procedures and science data. The size for
the high background subarray, (512$\times$512 pixels), is determined by the
dynamic range needed to image faint sources in the background glow of the Orion
Nebula region. The sizes for bright object subarrays are driven by the
saturation limits needed to observe known radial velocity planet host stars, 
but limited by the fact that going below a 72 $\times$ 64 subarray does not 
yield a significantly faster pixel clocking speed because of the overheads discussed above. 
The size of the slitless prism subarray is determined by the number of
pixels needed to cover the 5-12 \micron{} LRS spectrum in the dispersion
direction and to provide adequate sky observations for background
subtraction in the spatial direction.

\section{Performance}

\subsection{Electro-optical properties}

The basic performance of the MIRI detector arrays is described by Ressler et
al. (2008) and summarized in Table 2. That paper describes the measurements
of quantum efficiency, response vs detector bias voltage, read noise, well
depth, and a number of other parameters that it is not necessary to update.
We describe below areas where significant changes in methodology have
occurred since publication of that paper.

As described in Table 2 and in Paper VII, there are two slightly different detector architectures 
in the MIRI arrays, termed the baseline and the contingency designs. 
The contingency array (used in the short wavelength arm of the MRS) trades a
lower absorption efficiency (due to lower doping and a thinner IR-active
layer) for reduced dark current. The baseline array type is used in both the
imager and the long wavelength arm of the MRS, but with the 6$\mu$m AR
coating for the former and the 16$\mu$m one for the latter.

\subsubsection{Pixel gain}

The pixel gain, defined as the digital numbers out of the electronics per
electron placed on the integrating node of an array amplifier, is a critical
parameter for interpretation of many aspects of array and instrument
behavior. We have measured the gain by injecting a test voltage into
V$_{dduc}$ and tracking V$_{detcom}$ with that variation (to keep the
detector bias constant at 2.0V). Please refer to Figure 4 for the roles of
these voltages. By monitoring the FET output as a function of V$_{detcom}$,
we find a net system gain of 38300 DN / V, where V is the voltage into the
integrating node. To convert this measurement to a pixel gain requires the
capacitance of the integrating node. The value for the input to the MIRI
multiplexer has been measured to be 28.5 fF (McMurtry 2005), to which we
need to add contributions from a detector pixel (1.1 fF from basic physics)
and from the bump bond interconnects between the detectors and their
amplifiers (about 4 fF, e.g. Moore 2005). For the resulting estimate of 33.6
fF, the gain is $\sim$ 5.5 electrons/DN. That is, a signal of 5.5 electrons
from a detector results in one DN change in the FPE output.

\subsubsection{Dark current}

Dark currents in the flight detector arrays were measured as part of the
flight model instrument test campaign. The contamination control cover (CCC; see Paper II)
was closed to make the instrument interior as dark as possible (although as
always one can measure only upper limits to the true dark current, given the
possibility of photon leaks). In processing the data, the first and last
frames of an integration ramp were rejected, to circumvent the effects of
the reset anomaly and last-frame effect (see below). Dark currents were then
determined by the slopes of the integration ramps over a 100$\times$100 pixel
region selected to avoid bad pixels. The slope calculations included all
exposures in a test run; the effects of settling of the detector output
artificially elevate the apparent dark current, again making the results
upper limits.

\subsubsection{Imaging properties}

The response of the arrays is uniform, with pixel-to-pixel variations of no
more than 3{\%} rms. The best arrays have a small proportion of inoperative
pixels (either hot or dead), of order 0.1{\%}.

The imaging properties of MIRI were measured multiple times during the
buildup of the instrument modules and then in the test of the flight model
prior to delivery, and finally in the Integrated Science Instrument Module
(ISIM) test post delivery. The most precise of these measurements in terms
of the array performance were conducted at CEA with just the imager, and
illuminated by a source outside the cryostat with extensive filtering to
control the background emission (e.g., Ronayette et al. 2010). The point
spread function (PSF) measurements utilized a micro-stepping strategy so
that many positions of the source were recorded, on centers smaller than the
pixel pitch of the MIRI array. These measurements were then converted to a
high-resolution PSF image. The images at the longer wavelengths are as
expected (Paper VII), showing excellent imaging properties from the
array with only a low level of crosstalk. 
We have measured the pixel-to-pixel crosstalk in a number of ways
(Finger et al. 2005; Regan {\&} Bergeron 2012; Rieke {\&} Morrison 2012),
including autocorrelation, analysis of cosmic ray hits, and of hot pixels.
All measurements indicate a level close to 3{\%} for the crosstalk to the
four adjacent pixels around one receiving signal. A plausible cause of this
behavior is interpixel capacitance, although there may be secondary
contributions from electron diffusion and optical effects (Rieke {\&}
Morrison 2012), In addition, at 5.6 $\mu $m there is an additional
cross-like imaging artifact, discussed in more detail in Paper VII

\subsection{Non-ideal behavior}

In common with most infrared arrays, the MIRI arrays show a variety of
non-ideal behaviors, the most important of which are described below. With
the exception of the last frame effect, previous generations of Si:As IBC
detector arrays show all of the effects seen in the MIRI ones. The goal of
the ongoing pipeline development is largely to mitigate these effects,
making use both of experience with previous similar detector arrays and through 
a test campaign with similar arrays and electronics, with analysis of the
results by the MIRI pipeline development team. The primary areas of interest
are listed below.

\subsubsection{Reset anomaly}

With the MIRI arrays, the first few samples starting an integration after a
reset do not fall on the expected linear accumulation of signal. This
behavior is seen in virtually all types of infrared arrays, both of Si:X IBC
type (e.g., Gordon et al. 2004; Hora et al. 2004), and more generally (e.g.,
Rauscher et al. 2007; Rieke 2007). The reset anomaly can be removed largely
(or perhaps entirely) by subtracting from each ramp under signal, 
a correction generated from the behavior of that ramp in the
dark. The approach to
doing this has to be robust to changing "dark" slopes due to slow detector
settling (e.g., after a change in background level or following events such
as a thermal anneal or just turning the detector on). It has been found that
subtracting the median slope of a series of dark integration ramps before
subtracting from a data ramp can provide a good correction that is largely
immune to these effects.

\subsubsection{Last frame effect}

The array is reset sequentially by row pairs (Section 3.1). The last frame
of an integration ramp on a given pixel is influenced by signal coupled
through the reset of the adjacent row pair. The result is that the odd and
even rows both show anomalous offsets in the last read on an integration
ramp.

\subsubsection{Droop}

The Rockwell/Boeing North America Si:X IBC arrays used in Spitzer IRS and
MIPS had outputs for all the individual pixels that included a component
proportional to the total signal received by the array (with a
proportionality constant of 0.3 to 0.4). This phenomenon was termed "droop"
(van Cleve et al. 1995). Similar behavior was exhibited by the WISE Si:As
IBC arrays, but the effect is greatly reduced in the IRAC and MIRI arrays. A
plausible explanation is that the latter two arrays use the Raytheon
cryogenic readout manufacturing process that heavily dopes the multiplexer substrate to
within a few microns of the circuitry to improve the performance of the
ground plane in limiting long term drifts and other undesirable types of
behavior. It appears that it may not be necessary to make corrections for
droop in the MIRI pipeline.

\subsubsection{Drifts}

The MIRI arrays are subject to slow output drifts with amplitudes roughly
proportional to the signal level (e.g., the drifts are greatly reduced when
the arrays are in the dark). A simple description of this behavior, along
with a prescription for removing it, is that an observing strategy that
includes dithers so that images can be subtracted while retaining the source
signals appears to be able to remove the effects of the drifts, so long as
the dithers are performed at least every five minutes. There are some other
aspects not captured in this simple picture; for example, the amplifier zero points
also drift in a way that can affect the linearity corrections. This behavior is
similar to that for the Spitzer Si:X IBC arrays, both from Raytheon and
Boeing. For example, it was necessary with IRAC to dither every 7 minutes 
to generate optimum self-calibrated flat fields,
while the IRS allowed integrations only up to 8 minutes before moving the
source on the slit (from the IRAC and IRS instrument handbooks, {\it
  Spitzer} Science Center, 2011, 2013).

Although dithering is an acceptable strategy for many MIRI operational
modes, for coronagraphy, planetary transit observations, and some others it
is not a viable option. The MIRI array reference pixels do not follow the
drifts accurately enough to remove them noiselessly. 
The decay of latent sources is likely one component of the cause of the drifts.
Therefore, learning how to deal with latent images  as well as other causes of
the drifts is an open issue in the development of the MIRI pipeline.

\subsubsection{Multiplexer Glow}

The MIRI multiplexer can contribute significant levels of glow to the
low-level signals from, for example, the MIRI spectrometers. A primary
source of this emission is both the column and row shift registers, where it
is associated with the forward biasing of the n-FET substrates of the shift
register MOSFETs. Adjustment of the bias potentials of the p-wells in which
these MOSFETs reside can minimize their glow. Further reduction can be
achieved by control of the potentials for the rails of each of the shift
register clocks. The upgraded electronics boards allow
greater ability to optimize these settings than with the original MIRI FPE.

\subsubsection{Latent Images}

Bright sources leave latents on the MIRI arrays, typically at a level of
about 1{\%} immediately after the source has been removed. The decay of
these images shows multiple time constants, suggesting that there are a
number of mechanisms that contribute to the effect (Paper VII). Further
characterization of this complex behavior is needed to determine ways to
correct it in the MIRI pipeline.

\section{Pipeline development}

The non-ideal aspects of the MIRI detector behavior are not new, having been
seen in all previous examples of similar detector arrays. However, producing
well-calibrated data from the instrument requires that these anomalies be
understood thoroughly. To do so, the MIRI team is carrying out a series of
test runs at JPL using flight-clone electronics and flight-like detector arrays to
study the performance in depth. The results from these tests are analyzed by
the extended MIRI team (Space Telescope Science Institute, European
Consortium, JPL, University of Arizona). This work, plus theoretical studies
of the detectors, is reported among the team and used to generate and
improve the algorithms for the data pipeline. The architecture of the pipeline
is described in Paper X and allows for adding the steps needed to optimize the detector performance as they are determined. 
We expect this effort to
continue virtually until launch of JWST.

\section{Acknowledgements}
The work presented is the effort of the entire MIRI team and the enthusiasm within the MIRI partnership is a significant factor in its success. MIRI draws on the scientific and technical expertise many organizations, as summarized in Papers I and II. 
A portion of this work was carried out at the Jet Propulsion Laboratory, California Institute of Technology, under a contract with the National Aeronautics and Space Administration.

In addition, Alan Hoffman and Peter Love were central to the development of the MIRI
arrays at RVS before their retirements. We also thank John Drab for
overseeing the completion of the arrays and George Domingo for his advice
and assistance throughout. Craig McCreight, Mark McKelvey, and Bob McMurray
led early work to demonstrate the large format Si:As IBC arrays. Thanks to
Hyung Cho and Johnny Melendez of JPL for many hours
invested in testing the SCAs and FPS system. The research described in this
paper was carried out at the Jet Propulsion Laboratory, California Institute
of Technology, under a contract with the National Aeronautics and Space
Administration.  Additional support was provided through NASA grant NNX13AD82G, 
and by the Centre Nationale D'Etudes Spatiales (CNES), UK Science and Technology Facilities
Council, and the UK Space Agency.


\clearpage

\begin{deluxetable}{lccc}
\tabletypesize{\footnotesize}
\tablecolumns{6}
\tablewidth{0pt}
\tablecaption{MIRI subarray properties}
\tablehead{\colhead{}           &
           \colhead{Size}             &
	 \colhead{}             &
 	\colhead{Frame time}             \\
						
	\colhead{Subarray}          &
	\colhead{columns by rows}          &
	\colhead{Starting position}          & 
	\colhead{(seconds)}             
 }
\startdata
FULL           & 1032 X 1024 & (1,1)    & 2.775  \\
BRIGHTSKY      &  512 X  512 & (457,51) & 0.865  \\
SUB256         &  256 X  256 & (413,51) & 0.300  \\
SUB128         &  136 X  128 & (1,889)  & 0.119  \\
SUB64          &   72 X   64 & (1,779)  & 0.085  \\
SLITLESSPRISM  &   72 X  416 & (1,529)  & 0.159  \\   
MASK1065   &  288 X  224 & (1,19)   & 0.240 \\
MASK1140   &  288 X  224 & (1,245)  & 0.240\\
MASK1550   &  288 X  224 & (1,467)  & 0.240  \\ 
MASKLYOT   &  320 X  304 & (1,717)  & 0.324 \\
\enddata
\end{deluxetable}

\clearpage

\begin{deluxetable}{lcc}
\tabletypesize{\footnotesize}
\tablecolumns{3}
\tablewidth{0pt}
\tablecaption{MIRI detector array properties}
\tablehead{\colhead{Parameter (units)}           &
           \colhead{Baseline array}             &
           \colhead{Contingency array}          }
\startdata
Format (pixels) & 1024 X 1024 & 1024 X 1024  \\
Pixel pitch ($\mu$m) & 25 & 25  \\
Arsenic concentration (cm$^{-3})$ & 7 $\times$ 10$^{17}$  & 5 $\times$ 10$^{17}$  \\
IR-active layer thickness ($\mu$m)  &  35  &  30  \\
AR coating peak ($\mu$m)   &     6 \& 16   &   6   \\
Read noise (e rms, Fowler-8)  &  $\sim$ 14  &   $\sim$ 14  \\   
Dark current (e/s)  &  $\sim$ 0.2     &  $\sim$ 0.07    \\
Quantum efficiency (\%) &  &  \\
~~~~~~~7 - 12$\mu$m$^a$   &   $\ge$ 55  &   $\ge$ 40   \\ 
~~~~~~~12 - 27$\mu$m    &  $\ge$ 60  &   $\ge$ 50  \\
Latent images$^b$ (\%)   &  $\sim$ 0.5  &    -    \\
Minimum integration time$^c$ (s)  & 2.78   &  2.78 \\
Full well (electrons) & $\sim$ 250,000 & $\sim$ 250,000 \\
Minimum detectable signal$^d$ mJy   & 0.00006    &  -    \\
Maximum signal$^e$ (mJy)   &  420   &  - \\
\enddata
\tablenotetext{a}{QE for baseline array applies to the imager version}
\tablenotetext{b}{3 minutes after removing signal}
\tablenotetext{c}{For full frame readout}
\tablenotetext{d}{3-$\sigma$, 10,000 seconds, point source; this value and the following one are provided as examples of the net dynamic range of the FPS.}
\tablenotetext{e}{For 64 X 64 subarray, 5.6$\mu$m imaging}
\end{deluxetable}

\clearpage

\clearpage

\clearpage

\clearpage

\clearpage

\clearpage

\clearpage

\clearpage

\clearpage

\end{document}